\documentstyle[12pt]{article}
\def\d{\partial}

\def\ln{{\rm ln}}

\def\0{\nonumber}
\def\ee{\end{eqnarray}}     
\def\be{\begin{eqnarray}}
\def\ba{\begin{array}}          
\def\ea{\end{array}}

\begin{document}
\begin{flushright}
{SISSA 85/2002/EP}\\
{hep-th/0211283}
\end{flushright}

\begin{center}
{\LARGE {\bf  Integrable Structures in String Field Theory}}

\vskip 1cm

{\large L. Bonora$^a$ and A.S. Sorin$^b$}
{}~\\
\quad \\
{\em ~$~^{a}$International School for Advanced Studies (SISSA/ISAS),}\\
{\em Via Beirut 2, 34014 Trieste, Italy and INFN, Sezione di Trieste}\\
 {\tt bonora@sissa.it}
{}~\\
\quad \\
{\em ~$~^{b}$Bogoliubov Laboratory of Theoretical Physics,}\\
{\em Joint Institute for Nuclear Research (JINR),}
\\
{\em 141980, Dubna, Moscow Region, Russia}\\
{\tt sorin@thsun1.jinr.ru}

\end{center}

\vskip 2cm {\bf Abstract.} {We give a simple proof that the
Neumann coefficients of surface states in Witten's SFT satisfy the
Hirota equations for dispersionless KP hierarchy. In a similar way
we show that the Neumann coefficients for the three string vertex
in the same theory obey the Hirota equations of the dispersionless
Toda Lattice hierarchy. We conjecture that the full
(dispersive) Toda Lattice hierachy and, even more attractively
a two--matrix model, may underlie open SFT.}

\vskip 1cm

\section{Introduction}

A remarkable integrable structure underlies Witten's String Field
Theory (SFT), \cite{W1}. It manifests itself in the Neumann
coefficients, which can be shown to satisfy the Hirota bilinear
equations. This is true for the Neumann coefficients of the surface
states such as the sliver and the butterfly states,
\cite{KP,RZ,GRSZ1,GRSZ2}. In this case the
relevant integrable hierarchy is the dispersionless KP (dKP) hierarchy.
But, what is more important, it is also true for the Neumann coefficients
that define the three string vertex, a structure which is at the core
of Witten's SFT since it describes the basic interaction and the star
product. At the tree level in this case the relevant integrable hierarchy
is the dispersionless Toda lattice (dTL) hierarchy.

Following an inspiring suggestion of \cite{BR}, in this paper we
show in a straightforward way that the Neumann coefficients for
surface states satisfy the Hirota equations for the dKP hierarchy
and that the three string Neumann coefficients satisfy the Hirota
bilinear equations for the dTL hierarchy. Since the dKP hierarchy
is a subcase of the dTL hierarchy, it is evident that SFT is based
(at the tree level) on the latter.

These results are simple verifications that some identities hold
for Neumann coefficients in SFT, but they may have important
implications. For it is naturally tempting to conjecture that
the full SFT is based on the (dispersive) TL hierarchy. And
since the latter is the hierarchy underlying two--matrix models,
one is lead to conjecture that at the basis of SFT there may lie a
two--matrix model. So the discovery of the integrable structure of
Witten's string field theory opens the way to a new (old) exciting
field of research.

The paper is organized as follows. In section 2 we introduce the
Hirota equations for dKP and dTL hierarchies. In section 3 we
introduce new compact formulas for Neumann coefficients for
surface states and verify that they satisfy the Hirota equations
for the dKP hierarchy. In section 4 we introduce generating
functions for the Neumann coefficients of the three string vertex
and verify that they satisfy the bilinear Hirota equations for the
dTL hierarchy. Section 5 is devoted to comments and future
prospects.

\section{dKP and dTL Hirota equations}

The essence of integrable hierarchies is contained in the Hirota
bilinear equations \cite{TL}. We present them here for the dKP and
dTL hierarchies \cite{K,TT} . To this end let us introduce the
flow parameters $t_0, t_k, \bar t_k$ with $k= 1,2,..,\infty$ and
the differential operators
 \be
 D(z) = \sum_{k=1}^\infty \frac 1{kz^k} \frac {\d
}{\d t_k}, \quad\quad \bar D(\bar z) = \sum_{k=1}^\infty \frac 1{k
{\bar z}^k} \frac {\d }{\d {\bar t}_k}.\label{DDbar}
 \ee
Next we need the $\tau$--function (free energy) $F= F(\{t_0,t_k,
\bar t_k\})$. For the dKP $F$ depends only on $\{t_k\}$, $k\geq
1$. Then the Hirota equations for the dKP hierarchy can be written
in the following compact elegant form (for more details, see
\cite{Zabrodin, CD} and references therein):
 \be (z_1-z_2
)e^{D(z_1)D(z_2)F} + (z_2-z_3)e^{D(z_2)D(z_3)F} + (z_3-z_1
)e^{D(z_3)D(z_1)F}=0. \label{HdKP}
 \ee
The dispersionless Toda Lattice hierarchy depends instead on all
the parameters $\{t_0,t_k,\bar t_k\}$. The Hirota equations can
still be written in compact form
 \be
&&(z_1-z_2) e^{D(z_1)D(z_2)F} = z_1 e^{-\d_{t_0}D(z_1)F}
- z_2 e^{-\d_{t_0}D(z_2)F}, \label{HdTL1}\\
&& z_1\bar z_2 \left( 1 - e^{-D(z_1)\bar D(\bar z_2)F}\right)=
e^{\d_{t_0}(\d_{t_0} + D(z_1)+\bar D(\bar z_2))F} \label{HdTL2}
 \ee
and their bar-counterparts.

Our task in this paper is to identify the exponents in the above
equations (derivatives of $F$) with suitable generating functions
of Neumann coefficients and verify that the relevant Hirota
equations are identically satisfied.

\section{Neumann coefficient for surface states}

Neumann coefficients for surface states were defined in
\cite{LPPII}. Given a conformal map $f(z)$ from the upper
semidisk, they are defined by
 \be
N^f_{nm} = \frac 1{nm} \oint_0 \frac {dz}{2\pi i} \frac 1{z^n}
\oint_0 \frac {dw}{2\pi i} \frac 1{w^m} \d_z \d_w \ln \left( f(z)
- f(w)\right). \label{Nfnm}
 \ee
In the following we actually only need that $f(z)$ admits a Taylor
expansion around $z=0$, with $f'(0)\neq 0$. In (\ref{Nfnm}) the
integration contours are counterclockwise about the origin.

Now we define the generating function $N^f(z_1,z_2)$ for large
$z_1$ and $z_2$,
 \be
N^f(z_1,z_2) \equiv \sum_{n=1}^\infty \sum_{m=1}^\infty \frac 1
{z_1^n} \frac 1{z_2^m}\,N^f_{nm}. \label{genf}
 \ee
Using (\ref{Nfnm}), we make this definition more explicit as
follows
 \be
 N^f(z_1,z_2) &=& \oint_0 \frac {dw}{2\pi i} \oint_0 \frac
{d\zeta}{2\pi i} \,\sum_{n,m=1}^\infty\frac 1{n(z_1w)^n}\, \frac
1{m (z_2\zeta)^m}\, \d_w \d_\zeta
\ln \left(f(w)-f(\zeta)\right)\0\\
&=& \oint_0 \frac {dw}{2\pi i} \oint_0 \frac {d\zeta}{2\pi i} \,
\ln (1- \frac 1{z_1w}) \, \ln (1- \frac 1{z_2\zeta})\,\d_w
\d_\zeta
\ln \left(f(w)-f(\zeta)\right)\label{Nf1}\\
&=& \oint_0 \frac {dw}{2\pi i} \oint_0 \frac {d\zeta}{2\pi i} \,
\d_w \ln (1- \frac 1{z_1w})\, \d_\zeta \ln (1- \frac
1{z_2\zeta})\, \ln \left(f(w)-f(\zeta)\right)\0.
 \ee
The series converge for $|z_1 w|>1$ and $|z_2 \zeta|>1$.  The
integration contours are chosen in such a way as to surround the
logarithmic cuts. Next we insert the decomposition
 \be
\d_w \ln (1- \frac 1{z_1w})\, \d_\zeta \ln (1- \frac 1{z_2\zeta})=
\frac 1{w\zeta} - \frac 1w \frac 1{\zeta- z_2^{-1}} - \frac 1\zeta
\frac 1{w- z_1^{-1}} + \frac 1{w- z_1^{-1}} \frac 1{\zeta-
z_2^{-1}}. \label{Nf2}
 \ee

We see immediately that the term $\frac 1{w\zeta}$ gives rise to a
divergent term in (\ref{Nf1}). For this reason we regularize
$N^f(z_1,z_2)$ by subtracting
 \be
 \oint_0 \frac {dw}{2\pi i} \oint_0
\frac {d\zeta}{2\pi i} \, \d_w \ln (1- \frac 1{z_1w})\, \d_\zeta
\ln (1- \frac 1{z_2\zeta})\, \ln (w-\zeta)\0
 \ee
  from it. Repeating the derivation, inserting (\ref{Nf2}) and
performing the contour integrals, we get
 \be
N^f(z_1,z_2) &=& \oint_0 \frac {dw}{2\pi i} \oint_0 \frac
{d\zeta}{2\pi i} \sum_{n,m=1}^\infty \,\frac 1{n(z_1w)^n}\, \frac
1{m (z_2\zeta)^m}\, \d_w \d_\zeta
\ln \left(\frac{f(w)-f(\zeta)}{w-\zeta}\right)\0\\
&=& \ln \left( \frac {f'(0)}{z_1-z_2} \, \frac{f(z_1^{-1})-
f(z_2^{-1})} {(f(0) - f(z_2^{-1}))(f(z_1^{-1})-f(0))} \right).
\label{Nfgen}
 \ee

We stress that the subtracted term with $\ln(w-\zeta)$
does not change the value of the initial integral
because it corresponds to the identity mapping with obviously
trivial Neumann coefficients.

Now we identify $N^f(z_1,z_2)$ with $D(z_1)D(z_2)F$ in
eq.(\ref{HdKP}),
 \be
&&D(z_1)D(z_2) F = N^f(z_1,z_2), \label{ded111}
 \ee
and analogously for the other exponents. It is now elementary to
prove that (\ref{HdKP}) is identically satisfied. We simply remark
that $N^f(z_1,z_2)$ (\ref{Nfgen}) can be rewritten as
 \be N^f(z_1,z_2)= \ln
\left( \frac{f'(0)}{z_1-z_2} \, \left(\frac 1{(f(0) -
f(z_2^{-1}))}- \frac 1
{(f(0)-f(z_1^{-1}))}\right)\right),\label{Nfgen'}
 \ee
then plug this equation in (\ref{HdKP}) and use cyclicity.

The fact that $N^f(z_1,z_2)$ satisfies the Hirota equations means
that the corresponding Neumann coefficients $N^f_{nm}$
(\ref{genf}) satisfy all the identities implied by (\ref{HdKP}),
which can be obtained by expanding the latter in powers of
$z_1,z_2$ and $z_3$, and equating the corresponding coefficients.
The first few of these identites have been written down in
\cite{CD,BR} and will not be repeated here. Instead we would like
to point out that formula (\ref{Nfgen'}) is very efficient in
explicitly computing the relevant Neumann coefficients. A very
simple example is given by the so--called Nothing state,
\cite{GRSZ2}. In this case
 \be
f(z) = f_N(z)\equiv \frac {z}{1+z^2}. \label{noth0}
 \ee
 Plugging in this formula one
can obtain the following expression for the generating function
$N^{f_{N}}(z_1,z_2)$ of the Neumann coefficients $N^{f_N}_{nm}$
of the Nothing state
 \be N^{f_{N}}(z_1,z_2)= \ln \left( 1- \frac{1}{z_1 z_2}
\right).\label{noth1}
 \ee
Then, the Neumann coefficients can easily be extracted by
expanding $N^{f_{N}}(z_1,z_2)$ using (\ref{genf})
 \be N^{f_N}(z_1,z_2) \equiv
-\sum_{n=1}^\infty \sum_{m=1}^\infty \frac 1 {z_1^n} \frac
1{z_2^m}\,\frac{1}{\sqrt{nm}}~\delta_{n,m}, \quad
N^{f_N}_{nm}=-\frac{1}{\sqrt{nm}}~\delta_{n,m} \label{noth2}
 \ee
and, up to normalization, reproduce correctly the corresponding
coefficients given in \cite{GRSZ2}.

As a second example, we mention the series of states
characterizing by \cite{GRSZ2}
 \be
 &&f_{B_n}(z) =  \exp \Bigl(
-{z^{n+1}\over n}{\partial\over\partial z} \Bigr)\, z ={z \over
(1+ z^n)^{1/n}} \, \quad n=1,2, \ldots \quad . \label{butt1}
 \ee
The state at $n=2$ corresponds to the so--called Butterfly state,
while the states with other $n$ are generalizations of the latter.
Plugging (\ref{butt1}) into eq. (\ref{Nfgen'}) we obtain the
corresponding generating functions $N^{f_{B_n}}(z_1,z_2)$ of
Neumann coefficients for these states
 \be
N^{f_{B_n}}(z_1,z_2)= \ln \left( \frac{(1+ z^n_1)^{1/n}-(1+
z^n_2)^{1/n}}{z_1-z_2} \right).\label{butt2}
 \ee
The Neumann coefficients $N^{f_{B_2}}_{nm}$ for the Butterfly
state were calculated explicitly in \cite{GRSZ2}, and it is a
simple exercise to verify that they are reproduced from eq.
(\ref{butt2}) at $n=2$. Let us remark that, actually, the method,
developed in \cite{GRSZ2} to evaluate $N^{f_{B_2}}_{nm}$, can also
be used to evaluate the Neumann coefficients $N^{f_{B_n}}_{nm}$
for any $n$, and we have explicitly verified that they are indeed
reproduced by our simple formula (\ref{butt2}).

\section{Neumann coefficients for three string vertex}

The three string vertex \cite{W1,GJ1,GJ2} of the Open String Field
Theory is given by
 \be
|V_3\rangle= \int d^{26}p_{(1)}d^{26}p_{(2)}d^{26}p_{(3)}
\delta^{26}(p_{(1)}+p_{(2)}+p_{(3)})\,{\rm exp}(-E)\,
|0,p\rangle_{123}\label{V3} \ee where \be E=
\sum_{r,s=1}^3\left(\frac 12 \sum_{m,n\geq 1}\eta_{\mu\nu}
a_m^{(r)\mu\dagger}V_{mn}^{rs} a_n^{(s)\nu\dagger} + \sum_{n\geq
1}\eta_{\mu\nu}p_{(r)}^{\mu}V_{0n}^{rs} a_n^{(s)\nu\dagger} +\frac
12 \eta_{\mu\nu}p_{(r)}^{\mu}V_{00}^{rs} p_{(s)}^\nu\right).
\label{E}
 \ee
Summation over the Lorentz indices $\mu,\nu=0,\ldots,25$ is
understood and $\eta_{\mu \nu}$ denotes the flat Lorentz metric.
The operators $ a_m^{(r)\mu},a_m^{(r)\mu\dagger}$ denote the
non--zero modes matter oscillators of the $r$-th string
($r=1,2,3$). $p_{(r)}^\mu$ are the center of mass momenta of the
$r$-th string. $|p_{(r)}\rangle$ is annihilated by the
annihilation operators $a_m^{(r)\mu}$ and is eigenstate of the
momentum operator $\hat p_{(r)}^\mu$ with eigenvalue
$p_{(r)}^\mu$.

The Neumann coefficients $V_{mn}^{rs},~V_{0n}^{rs}$ and
$V_{00}^{rs}$ were explicitly computed in \cite{GJ1}. For
definiteness, we refer to the coefficients contained in Appendix A
of \cite{RSZ2} as the {\it standard} ones.

Our purpose in this section is to show that these coefficients too
obey the Hirota equations. The difference with the previous section is
that these Hirota equations are the ones characteristic of the
dispersionless Toda Lattice hierarchy. We recall that in the
definition of $V_{0n}^{rs}$ there is a gauge freedom, since, due
to momentum conservation we are allowed to make the redefinition
 \be
 V_{0n}^{rs} \to V_{0n}^{rs} + A_n^s. \label{gauge}
 \ee
We will see below that
the Hirota equations involve only gauge invariant combinations of
them.

The strategy is the same as in the previous section. We first
define a more efficient method to calculate the Neumann
coefficient, by introducing suitable generating functions. A basic
difference with the previous section is that in this problem,
instead of one, we have  three functions involved
 \be
&&f_1(z_1) = e^{\frac{2\pi i}3}
\left(\frac {1+ i z_1}{1-iz_1}\right)^{\frac 23}\0 ,\\
&& f_2(z_2) = \left(\frac {1+ i z_2}{1-iz_2}\right)^{\frac 23},
\label{fi}\\
&& f_3(z_3) = e^{-\frac{2\pi i}3} \left(\frac {1+ i
z_3}{1-iz_3}\right)^{\frac 23}\0  .
 \ee
We define
 \be
&&N^{rs}(z_1,z_2) = \sum_{n,m=1}^\infty
\frac 1{z_1^n} \frac 1{z_2^m}
N^{rs}_{nm},\label{Nrs}\\
&& N_0^{rs}(z) = \sum_{n=1}^\infty \frac 1{z^n} N^{rs}_{0n}.
\label{N0rs}
 \ee
 Now, we insert in these definitions
the formulas for $N^{rs}_{nm}$ and $N^{rs}_{0n}$ given by
\cite{LPPI}, and, as in the previous section, we extract the
corresponding generating functions. The task is easy for $N^{rr}$,
because we can limit ourselves to taking the formula of the
previous section. In other words we have
 \be
N^{rr}(z_1,z_2) = \ln \left( \frac {f_r'(0)}{z_1-z_2} \,
\frac{f_r(z_1^{-1})- f_r(z_2^{-1})} {(f_r(0) -
f_r(z_2^{-1}))(f_r(z_1^{-1})-f_r(0))}\right). \label{Nrrgen}
 \ee
For $r\neq s$ we start again from \cite{LPPI}\footnote{Notice the
different sign with respect to \cite{LPPI}.}
 \be
N^{rs}_{nm} = -\frac 1{nm} \oint_0 \frac {dz}{2\pi i} \frac 1{z^n}
\oint_0 \frac {dw}{2\pi i} \frac 1{w^m} \d_z \d_w \ln \left(
f_r(z) - f_s(w)\right). \label{LPPN}
 \ee
Next we insert this formula in (\ref{Nrs}). We notice that, since
$r\neq s$ implies $f_r(0) \neq f_s(0)$, we do not need to do a
subtraction as in the previous section, so the method works
straight away. The result is
 \be
N^{rs}(z_1,z_2) = -\ln \left( \frac {(f_r(z_1^{-1}) - f_s
(z_2^{-1})) (f_r(0)- f_s(0))} { (f_r(0)- f_s (z_2^{-1}))
(f_r(z_1^{-1})- f_s(0))} \right). \label{Nrsgen}
 \ee

As for $N_0^{rs}(z)$ we start again from the formula that can be
found in \cite{LPPI}
 \be N_{0n}^{rs}=\frac 1m \oint_0 \frac
{dz}{2\pi i} \, \frac 1{z^n} \d_z \ln(f_r(0) -
f_s(z)),\label{LPPN0}
 \ee
and plug it into eq. (\ref{N0rs}). Once again we have to
distinguish two cases $r=s$ and $r\neq s$. The first case needs a
subtraction, the second does not. The result is
 \be N_0^{rr}(z) =
\ln \left( z\frac {f_r(0) - f_r(z^{-1})}{-f_r'(0)}
\right)\label{N0rrgen},
 \ee
while, for the case $r\neq s$, we obtain
 \be N_0^{rs}(z) = \ln
\left(\frac {f_r(0) - f_s(z^{-1})} {f_r(0)-f_s(0)} \right).
\label{N0rsgen}
 \ee

 From the above definitions it is immediate to see that
$$N^{rs}_{nm}= N^{sr}_{mn}.$$
Moreover, from the explicit form of the functions
$f_r$ (\ref{fi}), one can easily verify that
$$N^{r+1,s+1}=
N^{rs}, \quad N_0^{r+1,s+1}= N_0^{rs}$$ with $r$ and $s$ defined
mod 3.

Now, having all the generating functions at hand, we want to show
that they satisfy the Hirota equations (\ref{HdTL1}--\ref{HdTL2}).
To this end we proceed to identify the exponents of these
equations with the previous generating functions. When zero modes
are not involved things are straightforward. We have, for
instance,
 \be
&&D(z_1)D(z_2) F = N^{11}(z_1,z_2),\label{id1}\\
&& D(z_1)\bar D(\bar z_2) F = N^{12}(z_1,\bar z_2).\label{id2}
 \ee

The identification for the generating functions involving zero
modes is more complicated and can only be by trial and error,
because of the gauge freedom (\ref{gauge}) mentioned above. After
some attempts we came to the following identification:
 \be
&&\d_{t_0} D(z) F = N_0^{11}(z)- N_0^{21}(z),\label{id3}\\
&&\d_{t_0} \bar D(\bar z) F = N_0^{22}(\bar z)- N_0^{12}(\bar z).
\label{id4}
 \ee
We see that the combinations in the RHS are gauge invariant!

Now we can proceed to verify that our generating functions satisfy
the Hirota equations. Substituting (\ref{id1}) in the LHS of
(\ref{HdTL1}) and (\ref{id3}) on the RHS, it is trivial to verify
that the equations identically satisfied. Similarly, for eq.
(\ref{HdTL2}) we have to insert (\ref{id2}) in its LHS and
(\ref{id3}--\ref{id4}) in its RHS. However, in this case, we have
an indeterminate: $F_{00}:= \d_{t_0}^2 F$. After some elementary
algebra one finds that eq. (\ref{HdTL2}) is identically satisfied
provided
 \be
e^{F_{00}} = \frac {f_1'(0) f_2'(0)}{(f_1(0) - f_2(0))^2} = \frac
{16}{27}. \label{F00}
 \ee
We see that $F_{00}$ is identical to the standard coefficient
$-V_{00}^{rr}$.

The Neumann coefficients that one can extract by expanding the
generating function (\ref{Nrrgen}) and (\ref{Nrsgen}) coincide, up
to normalization, with the standard Neumann coefficients
$V_{nm}^{rs}$ (\ref{E}) \cite{RSZ2}. For instance, for $r\neq s$,
let us set $N^{12}(1/x,1/y)\equiv H(x,y)$. Then
 \be
V^{12}_{nm} = -\frac {\sqrt{nm}}{n!m!} \d_x^n \d_y^m H(0,0).
\label{V12}
 \ee
Here are some examples
 \be
V^{12}_{11} = -\frac {16}{27},\quad V^{12}_{12} = -V^{12}_{21}= -
\frac {32}{81} \sqrt{\frac 23}, \quad\quad V^{12}_{31} = \frac
{16}{{\sqrt 3}~3^5}, \quad\quad V^{12}_{23}= {\sqrt 2} \frac
{160}{3^6}\0 .
 \ee
More generally, on the basis of the above equations, we have the
following identifications among second derivatives of $F$,
standard Neumann coefficients and the Neumann coefficients
$N_{nm}^{rs}$:
 \be
&&F_{t_nt_m}\equiv\frac {\d}{\d t_n}
\frac {\d}{\d t_m}F\, = \,- \sqrt{nm} V^{11}_{nm}
\,= \, nm N_{nm}^{11},\label{ifentif1}\\
&&F_{t_n\bar t_m}\equiv \frac {\d}{\d t_n} \frac {\d}{\d \bar t_m}F\, =
\,- \sqrt{nm} V^{12}_{nm} \,= \, nm N_{nm}^{12},\label{ifentif2}\\
&&F_{t_0\bar t_n}\equiv\frac {\d}{\d t_0} \frac {\d}{\d \bar t_n}F\,
= \,\sqrt{\frac n2} \left(V^{12}_{0n} -V_{0n}^{22}\right)
\,= \, n \left(N_{0n}^{22}-N_{0n}^{12}\right),\label{ifentif3}\\
&&F_{t_0t_n}\equiv \frac {\d}{\d t_0} \frac {\d}{\d t_n}F\, = \,
\sqrt{\frac n2} \left(V^{21}_{0n} -V_{0n}^{11}\right) \,= \, n
\left(N_{0n}^{11}-N_{0n}^{21}\right).\label{ifentif4}
 \ee

The Neumann coefficients involving zero modes and  computed by
expanding (\ref{N0rrgen}) and (\ref{N0rsgen}), do not coincide in
general with the standard ones. This is no surprise since the
latter were calculated with a definite gauge choice.
As far as the $N_{0n}^{rs}$ coefficients are
concerned we see once again that the Hirota equations only
constrain gauge invariant combinations of them.

The Hirota equations  generate infinite many relations among the
coefficients $N^{rs}_{nm},~N^{rs}_{0n}$ and $F_{00}$, relations
which one can derive by expanding (\ref{HdTL1}--\ref{HdTL2}) in
powers of $z_1, z_2,\bar z_2$. For instance, from eq.
(\ref{HdTL1}) we get
 \be
&&F_{t_1t_1} = \frac 12 F_{t_0t_2} - \frac 12 (F_{t_0t_1})^2,\label{f1}\\
&&\frac 12 F_{t_1t_2} = \frac 13 F_{t_0t_3} +\frac 12 F_{t_0t_1} F_{t_0t_2}
- \frac 16 (F_{t_0t_1})^3,\label{f2}\\
 \ee
and, from (\ref{HdTL2}),
 \be
&&e^{F_{00}} = F_{t_1 \bar t_1},\label{f3}\\
&& e^{F_{00}}F_{t_0t_1} = \frac 12 F_{\bar t_1 t_2},\label{f4}
 \ee
and so on. All these equations are identically satisfied.

\section{Discussion}

{}From the above results there can be no remaining doubt that an
integrable structure underlies Open String Field Theory. At the
tree level this is the dTL hierarchy, which implies in particular
that the Neumann coefficients are powerfully constrained. This
result in itself may seem at first simply formal, since it was
already known how to calculate all these coefficients (but above
we have presented new effective formulas to compute them). However
this fact opens the way to new possibilities, which seemed to be
out of reach until now. For, on the one hand, the dispersionless
TL hierarchy is a limiting case of the (dispersive) Toda
Lattice Hierarchy (TLH) \cite{UT}. This hierarchy describes all
genus corrections to the dTL results. On the other hand, we know
that two--matrix models, \cite{2M}, are precisely realizations of
TLH, in the sense that any solution of two--matrix models satisfy
the Hirota equations of TLH. We recall that the Hirota equations
are the same for all two--matrix models, but any two--matrix model
is characterized in addition by specific coupling conditions (or
string equations). We do not actually know whether any solution of
TLH can be incorporated in a definite two--matrix model. However
this seems to be very likely, and, anyhow, we have indirect
evidence that two--matrix models have to do with string theory. In
the past specific two-matrix models were shown to describe
topological field theories chracterizing string vacua with central
charge $c<1$, \cite{DKK}, or $c=1$, \cite{BX2,BX3}. It is
therefore hard to refrain from conjecturing that a specific
two--matrix model underlies open SFT. If true, this would be a
powerful tool to extract exact higher genus results, and perhaps
also exact all genus results. In fact we recall that, while the
Hirota equations generate relations among the Neumann coefficients
(but do not completely determine them), a two--matrix model would
allow us, at least in principle, to fully determine them (see, for
instance, \cite{BX2,BX3} for some examples in an admittedly very
simple case).

Needless to say all we said above rings a bell: two--matrix models are
coming back via the Dijkgraaf--Vafa description of Yang--Mills
theory, see \cite{KM}. Perhaps the parallelism between the two
appearances of matrix models is not accidental.

Another line of research the results of this paper open up
involves their supersymmetric extension, which is of course
relevant for supersymmetric SFT. In this direction the relevant
object is likely to be the dispersionless  version of the
supersymmetric KP hierarchy proposed in \cite{LC}. Quite recently,
the $N=(1|1)$ supersymmetric dTL hierarchy was proposed in
\cite{KS} which can also be relevant for this problem. However we
do not have yet a full fledged version of the latter hierarchy,
e.g. its bilinear form (if any), so the research in this field
faces an additional obstacle and requires further investigations.

{}~

{}~

{\bf Acknowledgements} A.S. would like to thank SISSA for the kind
hospitality during the course of this work. This research was
supported by the Italian MIUR under the program ``Teoria dei
Campi, Superstringhe e Gravit\`a'' and by the RFBR-CNRS Grant No.
01-02-22005, RFBR-DFG Grant No. 02-02-04002, DFG Grant 436 RUS
113/669 and by the Heisenberg-Landau program.

\end{document}